\documentclass[aps,prl,amssymb,amsmath,twocolumn,showpacs]{revtex4}
\usepackage{times}
\usepackage{graphicx}
\usepackage{amsfonts}
\usepackage{latexsym}
\newcommand{\fig}[2]{\includegraphics[width=#1\columnwidth]{#2}}

\newlength{\bilderlength}

\newcommand{\tildegamma}{\gamma_L}
\renewcommand{\ol}{\overline}
\def\ep{{f}}
\newcommand{\vep}{\varepsilon}
\newcommand{\EQ}{\begin{equation}}
\newcommand{\EE}{\end{equation}}

\begin{document}

\title{\sffamily\bfseries\large The Freezing of Random RNA}
\author{Michael L\"assig$^{(1)}$ and Kay J\"org Wiese$^{(1,2)}$} 
\affiliation{$^{(1)}$Institut f\"ur theoretische Physik,
Universit\"at zu K\"oln, Z\"ulpicher Str.\ 77, 50937 K\"oln, Germany\\
$^{(2)}$Laboratoire de Physique Th\'eorique de l'Ecole Normale
Sup\'erieure, 24 rue Lhomond, 75005 Paris, France}
\date{\small\today}

\begin{abstract}

We study secondary structures of random RNA molecules by
means of a renormalized field theory based on an expansion
in the sequence disorder. We show that there is a
continuous phase transition from a {\em molten phase} at
higher temperatures to a low-temperature {\em glass phase}.
The primary freezing occurs above the critical temperature,
with local islands of stable folds forming within the
molten phase. The size of these islands defines the
correlation length of the transition. Our results include
critical exponents at the transition and in the glass
phase.
\end{abstract}
\maketitle

RNA has various important functions in the cell, it forms viral
genomes, and has been attributed a key role in the origin of life. RNA
molecules fold into {\em unique} compact configurations able to
perform catalytic functions, and they can act as templates for the
readout of sequence information. In this sense, they are nature's
compromise between DNA and proteins, which explains their likely role
in early evolution as well as their ubiquity in today's molecular
biology.  Typical RNA folds at room temperature consist of {\em stems}
(i.e., parts of the molecule forming a helical double strand
stabilized by Watson-Crick base pairing) linked by {\em loops} (i.e.,
stretches of unpaired monomers). These conformations are governed by
the energies of base pairing and backbone bending as well as by the
entropy of the loops; their statistical physics is quite
complicated. Yet, the problem is more tractable than protein folding
since the free energy of an RNA fold can be separated energetically
into that of its {\em secondary} and its {\em tertiary}
structure~\cite{Tinoco,Higgs.review}. Labeling the bases consecutively
along the backbone of the molecule from 1 to $L_0$, the secondary
structure of the fold is completely defined by the Watson-Crick pairs
$(s,t)$ ($1 \leq s < t \leq L_0$) subject to the constraint that
different pairs are either independent ($s < t < s' < t'$) or nested
($s < s' < t' < t$); see fig.~1. Thus, the secondary structure
contains purely ``topological'' information about the fold, which is
independent of the spatial configuration. Due to the constraint on
base pairings, secondary structures can always be represented by
planar diagrams as shown in fig.~1. The interactions satisfying this
constraint are often the dominant part of the free energy, so the
secondary structure of a fold can be determined
self-consistently. There are efficient algorithms to compute the exact
partition function of secondary structures for a given
sequence~\cite{McCaskill, HofackerAl}. Base pairings violating the
constraint (so-called {\em pseudoknots}) as well as additional
interactions between paired bases are important for the tertiary
structure of the molecule (i.e., the full spatial arrangement of stems
and loops) but they generate only small-scale rearrangements of the
secondary structure~\cite{Tinoco,Isambert}. While this separation of
energies is only approximate, it can be tuned experimentally by
varying salt concentrations in the solution~\cite{Tinoco}. Hence, a
theory of secondary structures is an important starting point for
understanding RNA conformations.

\begin{figure}
{\parbox{0.47\textwidth}{\leftline{{\raisebox{2.6cm}{\!\!\!(a)}\hspace*{-.4cm}
\fig{0.495}{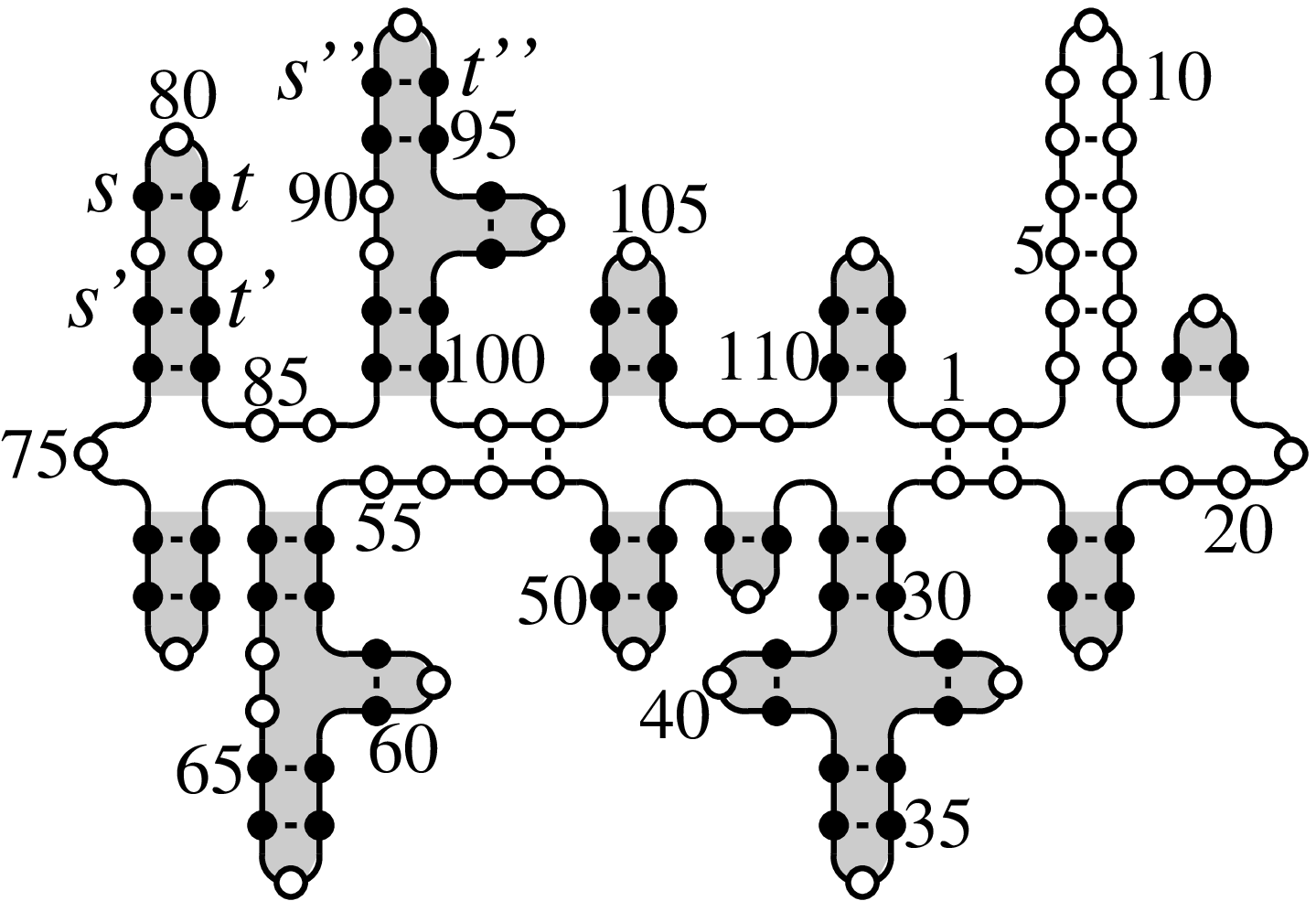} 
\hspace{-.1cm}
\fig{0.46}{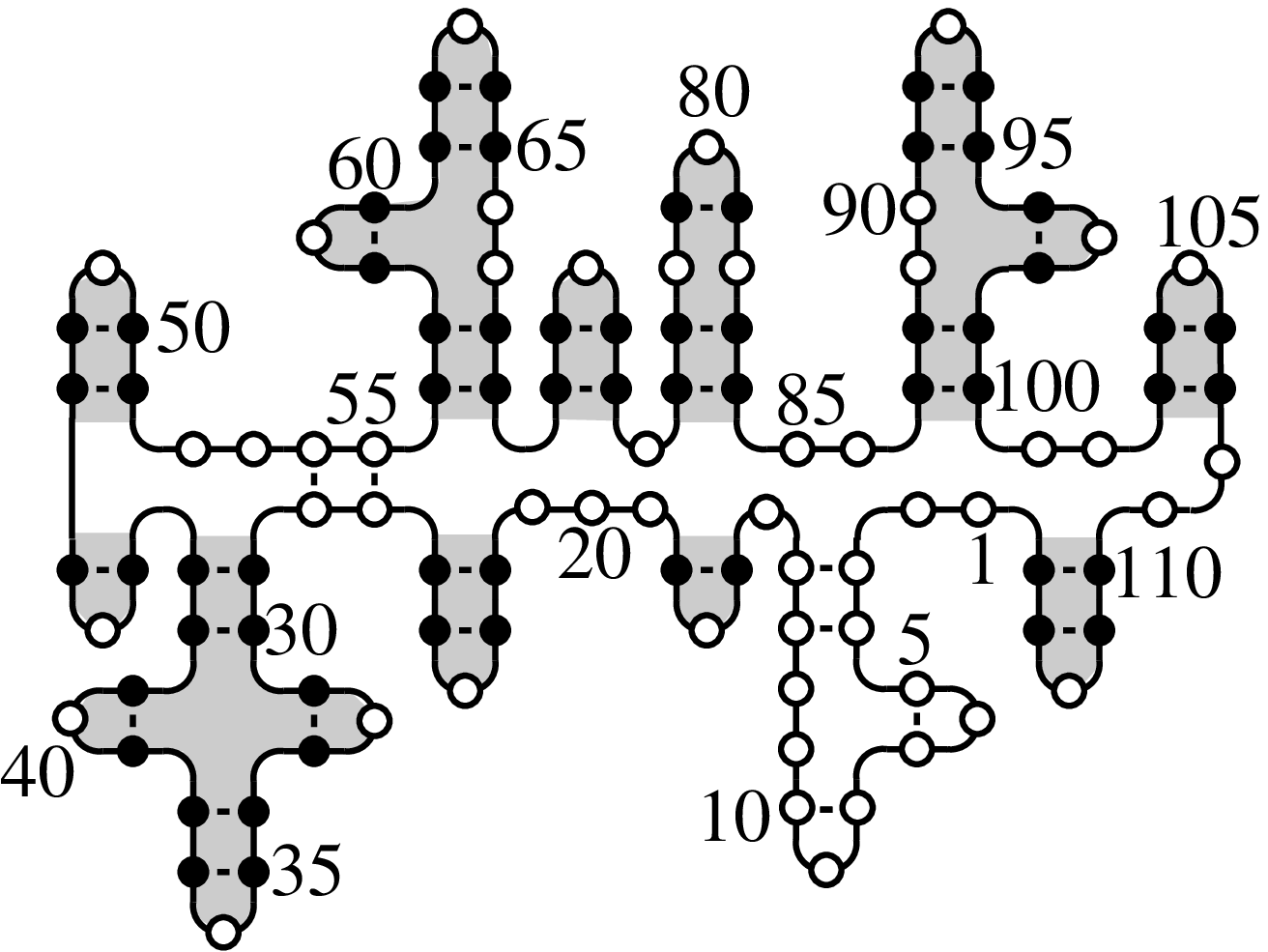}}}}} 
\vspace*{0.2cm}
\\ \hrule
\vspace*{0.2cm}
{\parbox{0.47\textwidth}{\leftline{{\raisebox{2.6cm}{\!\!(b)}\hspace*{-.35cm}\fig{0.495}{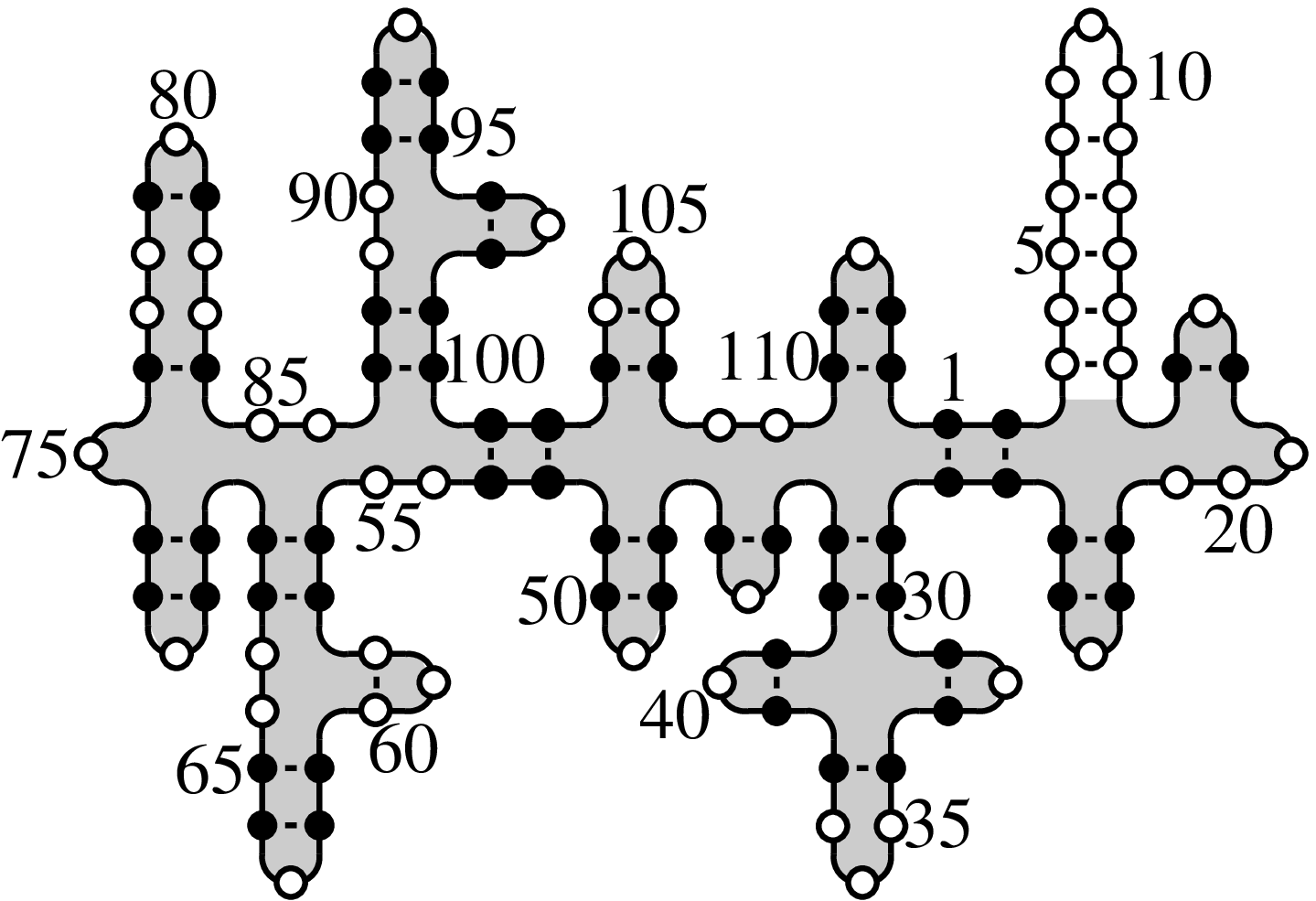} 
\hspace{-0.1cm}
\fig{0.495}{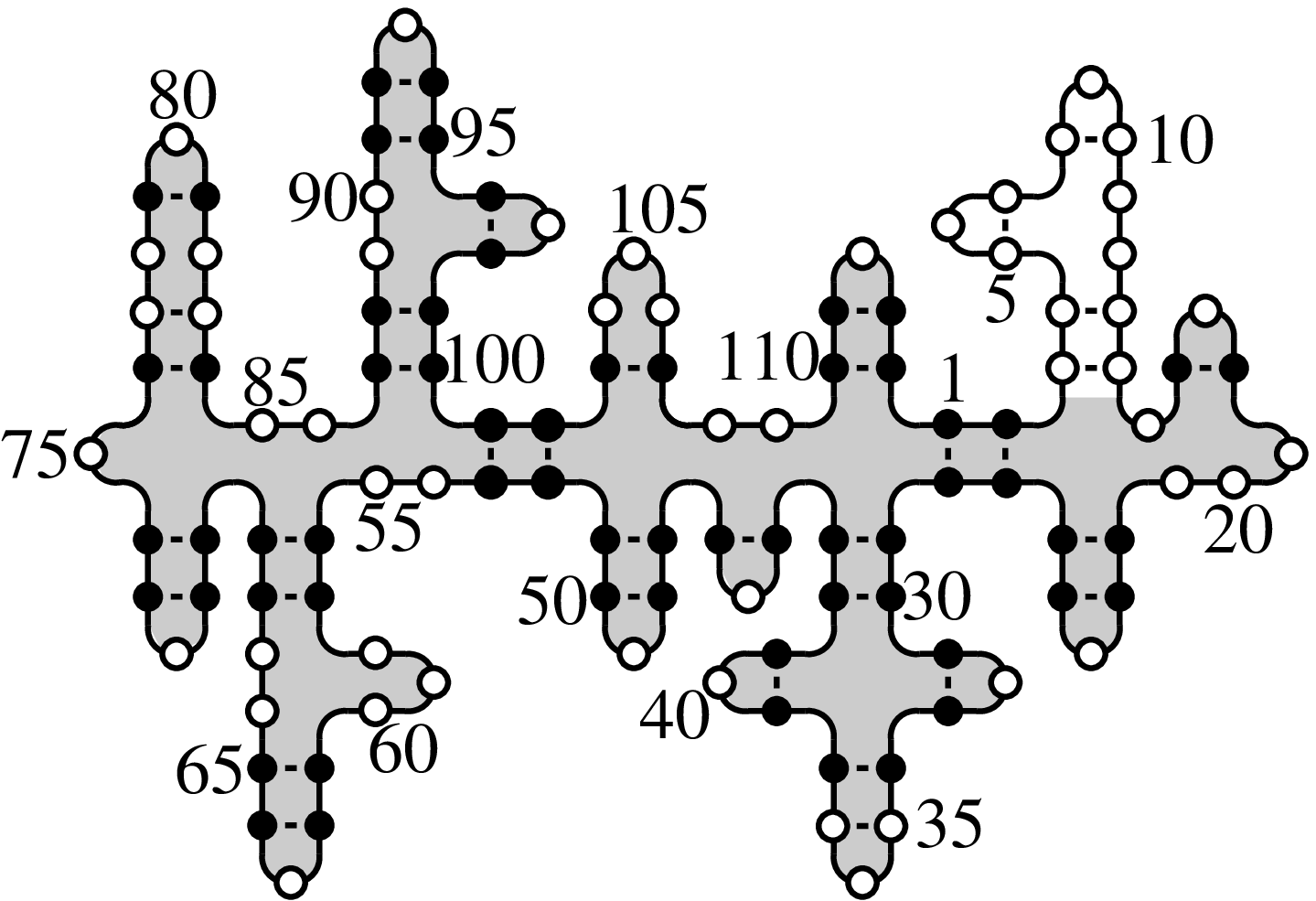}}}}}
\caption{Secondary structures of a random RNA molecule at
distant times.  Base pairings can be nested, such as
$(s,t)$ and $(s',t')$, or independent, such as $(s,t)$ and
$(s'',t'')$.  The pairing overlap is defined by the common
base pairings between the left and right configuration (the
corresponding bases are shown in black).  (a)~Above $T_c$,
the molecule contains conserved subfolds on scales up to
the correlation length $\xi$ (marked by
shading) and is molten on larger scales.  (b)~Below $T_c$,
the molecule is locked into its minimum energy structure on
all scales, up to rare fluctuations (unshaded).
}
\end{figure}

The simplest class of such molecules is {\em homopolymers}, where all
Watson-Crick pairs $(s,t)$ contribute an equal amount $\ep$ of free
energy. At room temperature, where $\ep$ is typically of order
$k_{\mathrm{B}}T$, homopolymers have a {\em molten phase} of compact
stem-loop folds.  The fold of an individual polymer in the molten
phase is not unique. It changes over time since thermal fluctuations
continuously build and undo its stems. The {\em pairing probability}
of two bases decays as a power law of their backbone
distance~\cite{deGennes},
$(t-s)^{-\rho_{0}}$, with $\rho_{0}= {3/2}$. 
In a {\em heteropolymer}, the energy of a Watson-Crick pair $(s,t)$
depends on the nucleotides at the backbone positions paired. An
important class is {\em random} heteropolymers. In biological systems,
such sequences result from evolution by {\em neutral}~\cite{Page}
mutations. For functional RNA, sequences and 
conformations are further modified by {\em
selection}, but
random sequences remain important as reference statistics. A
well-known analytical description of this case is to approximate the pairing
free energies $\eta(s,t)$ as independent Gaussian random variables
given by
\begin{equation}
  \ol{\eta (s,t)} = \ep, \;\;\; \ol{\eta(s,t) \eta(s',t')}
  -\ep^2 = \sigma^2 \delta (s - s') \delta (t - t'),
  \label{eta} 
\end{equation} 
where $\ep$ and $\sigma$ are of order $k_{\mathrm{B}}
T$~\cite{seq}. Free energy-estimates on the basis of this
model~\cite{BH99,BH02a,BH02b} and numerical
simulations~\cite{Higgs96,Pagnani2000,Hartmann01,BH02b,Marinari02,Krzakala02}
indicate that RNA random heteropolymers undergo a transition at a
critical temperature $T_c$ (about room temperature) from the molten
phase to a low-temperature {\em glass phase}. The nature of this phase
is controversial~\cite{BH02b,Marinari02,Krzakala02},
and the
numerical studies may suffer from significant finite-size
effects~\cite{Liu}. The two phases can be distinguished by
disorder-induced {\em replica} correlations. 
Replicas are simply two secondary structures 
at distant times -- i.e., drawn independently from the thermal
ensemble -- of the
same RNA molecule, i.e., the same disorder configuration
$\eta(s,t)$ as shown in fig.~1. Correlations between
replicas are defined by subsequent averaging over the
disorder distribution~(\ref{eta}). The 
arguments of \cite{BH02a,BH02b} for the {\em pairing overlap} (defined
as the joint probability of two bases being paired in both replicas)
suggest that replicas become independent at large backbone
distances in the molten phase but are essentially locked into a single
conformation in the glass phase.

In this letter, we develop a systematic field theory of random RNA
secondary structures.  This theory has two basic fields. The {\em
contact field} $\Phi (s,t)$ is defined to be 1 if the bases $s$ and
$t$ are paired and 0 otherwise. The {\em overlap field} between two
replicas $\alpha $ and $\beta$, defined as $\Psi_{\alpha \beta }(s,t) =
\Phi_\alpha (s,t)\Phi_\beta (s,t)$, describes correlations between the
replicas. By means of the {\em height field} $h(r) \equiv
\sum_{s=1}^{r} \sum_{t = r+1}^{L_0} \Phi(s,t)$, any secondary
structure can be mapped onto a random walk $h(r)$ ($r = 0, \dots,
L_0$) with step size $h(r) - h(r-1) = \pm 1$ and 
boundary conditions $h(0) = h(L_0) = 0$. This mapping relates
random RNA folds to the simpler problems of directed
polymers in a disordered medium~\cite{HuseHenley} and
Kardar-Parisi-Zhang surface
growth~\cite{KPZ,Lassig.review}. Generalizing existing field theoretic
approaches \cite{Lassig.KPZ,Wiese.KPZ,WieseHabil}, we derive
renormalization equations for the two fundamental variables of the
theory, the {\em disorder strength} and the {\em backbone length}.
The large-distance scaling of pairing probability and replica overlap are
given by the disorder-averaged expectation values 
\begin{equation}
\begin{array}{lll}
 \ol{\langle \Phi (s,t) \rangle } \sim (t - s)^{- \rho_0}, & \;
 \ol{\langle \Psi (s,t) \rangle } \sim (t - s)^{- \theta_0} & \;
 \mbox{($ T > T_c)$,}
 \\
 \ol{\langle \Phi (s,t) \rangle } \sim (t - s)^{- \rho^*}, & \;
 \ol{\langle \Psi (s,t) \rangle } \sim (t - s)^{- \theta^*} & \;
 \mbox{($ T \leq T_c)$.}  
 \end{array}    
 \label{PhiPsi}
\end{equation}
Here $\rho_0 = 3/2$ and $\theta_0 = 3$ are the known exponents of the molten
phase~\cite{deGennes,BH99}. At $T_c$, our renormalization group  gives first-order
values $\rho^* = \theta^* \approx 11/8$. As will become clear below, the equality
$\rho^* = \theta^*$ is an exact (though not rigorous) conclusion beyond first
order provided the renormalization group scenario sketched
in fig.~3 is qualitatively correct, i.e., the true
exponents are monotonic in $p$ at fixed $\vep$. This equality implies that  two
replicas are essentially locked into a single conformation already at the
transition. Hence,  the {\em leading} scaling is given by the minimum-energy
configuration for all temperatures $T \leq T_c$, i.e., the exponents $\theta^* =
\rho^*$ govern the glass phase as well. The  height fluctuations
\begin{equation} 
\ol{\left \langle (h(r) - h(r'))^2\right \rangle} \sim \left \{ 
\begin{array}{ll}
(r - r')^{2 \zeta_0} & \;\;\; \mbox{($T> T_c$)} 
\\
(r - r')^{2 \zeta^*} & \;\;\; \mbox{($T \leq T_c$)} 
\end{array}
\right.
\label{h}
\end{equation}
with $\zeta^* \approx 5/8$ are linked to the contact correlations by
the exact scaling relation $\zeta + \rho = 2$ in all phases, which
follows from the continuum representation of the $h$ field, $ h(r) =
\int_0^r {\rm d}s \int_r^{L_0} {\rm d} t \, \Phi(s,t)$~\cite{foot1},
and has been obtained previously in a closely related context~\cite{BL9602}. 
These exponents agree well with the numerical values
$\zeta_{\mathrm{glass}} = 0.65$~\cite{BH02b,Krzakala02} and
$\rho_{\mathrm{glass}} = 1.3(4)$~\cite{Krzakala02,Bundschuh.pc} for
$T=0$.

Our results show that the glass transition is of second order. A singular 
length scale 
\begin{equation}
\xi \sim |T - T_c|^{-\nu^*}, 
\label{xi} 
\end{equation}
whose exponent $\nu^* = 1/(2 - \theta^*) \approx 8/5$ is determined by
hyperscaling, describes the crossover scaling above and below the critical 
point. The resulting freezing scenario of random RNA molecules  is quite
intricate. It is illustrated in fig.~1, where we show snapshots of the same 
molecule at two distant times for
two different temperatures. Above $T_c$, the correlations (\ref{PhiPsi}),
(\ref{h}) scale with their critical
exponents $\rho^*, \theta^*,  \zeta^*$ up to backbone distances 
$(t - s)$ resp.~$|r - r'|$ of order $\xi$.
Hence, an RNA fold has essentially frozen ``islands'' of size $\xi$ (i.e., its
replicas are locked) but is molten on larger scales (its replicas become
independent), see fig.~1(a).  As $T$ approaches $T_c$ from above, the replica
correlation length  $\xi$ increases according to (\ref{xi}),  and the turnover
time between conformations by thermal fluctuations grows. We call this process
{\em primary freezing}. At criticality, there is still a power law distribution
of {\em rare thermal fluctuations} as discussed below. Lowering the temperature
below $T_c$, the correlation length decreases again and even these rare
fluctuations are removed from larger to smaller scales; this is called {\em
secondary freezing}, see fig.~1(b). 

To derive our renormalization group, we write the secondary structure partition
function of a given heteropolymer as a sum over the
contact field configurations, 
\begin{equation} {\cal Z}[\eta] =
 \sum_\Phi \exp [-\beta \sum_{1\leq s < t \leq L_0} \eta (s,t) \Phi(s,t)],
\end{equation} 
and study the disorder-averaged free energy $\ol {\cal F} = -
\beta^{-1} {\rm Tr}_\eta \log {\cal Z}[\eta]$ obtained from the distribution
(\ref{eta}).  In the replica formalism, this leads to a system of $p$
interacting homopolymers, 
 ${\cal Z}^{(p)} = \sum_{\Phi_1,\dots, \Phi_p} \exp ( - \beta
 {\cal H}^{(p)})$,
whose Hamiltonian~\cite{BH02a,BH02b}
\begin{equation}
 {\cal H}^{(p)} = {\ep_0} \sum_{\alpha}\sum_{s < t} \Phi_\alpha(s,t) 
      - \frac{g_0}{2} \sum_{\alpha \neq \beta} 
      \sum_{s < t}\Psi_{\alpha \beta} (s,t)
\end{equation} is given in terms of the contact fields $\Phi_\alpha$
($1\leq \alpha \leq p$) and the overlap fields $\Psi_{\alpha \beta}$
($1 \leq \alpha, \beta \leq p$, $\alpha \neq \beta$) with the coupling
constants $\ep_0 = \ep - \beta \sigma^2$ and $g_0 = \beta \sigma^2$.
The renormalization of this theory is based on analytic continuation
in the homopolymer exponent $\rho_0$, 
or equivalently, in
the scaling dimension $\vep := 2 \rho_0-2 $ of the coupling constant
$g_0$~\cite{freefield}.  In the limit $p \to 0$, the free energy
${\cal F}^{(p)} = - \beta^{-1} \log {\cal Z}^{(p)}$ reproduces that of the
random system, $\ol {\cal F} = \lim_{p \to 0} {\cal F}^{(p)}/p$.

The noninteracting theory ($g_0 = 0$) describes homopolymers
in the molten phase and is exactly solvable in the 
continuum limit, i.e., for molecules of backbone length 
$L_0 \gg 1$. The free energy for closed rings
is
${\cal F}_0 = p \, \rho_0 \log L_0$~\cite{deGennes,BH99}. 
The correlation function of $N$ contact fields
$\Phi_\alpha(s_i,t_i)$ 
describes constrained configurations of the molecule with
$N$ fixed pairings $(s_i, t_i)$ ($i = 1, \dots, N$). These 
pairings generate $N+1$ subrings 
of backbone lengths 
$\ell_1, \dots, \ell_N, \ell_{N+1} = L_0 - \sum_{j=1}^N
\ell_j$. Since the secondary structure fluctuations in the 
subrings are independent, 
this correlation takes the factorized form 
\begin{equation}
\langle \Phi_\alpha (s_1,t_1) \dots \Phi_\alpha (s_N, t_N) \rangle_0 = 
\frac{\ell_1^{-\rho_0} \dots 
      \ell_{N+1}^{-\rho_0}}{L_0^{-\rho_0}}.
\end{equation}
Overlap correlations factorize further into the contributions
of the single replicas upon insertion of the definition
$\Psi_{\alpha \beta}(s_i,t_i) = 
   \Phi_{\alpha}(s_i,t_i) \Phi_{\beta}(s_i,t_i)$.

In the presence of interactions, we write the free energy 
as a perturbation series, 
\begin{eqnarray}  
\label{fpert}
   \lefteqn{{\cal F}(g_0,L_0) = {\cal F}_0 
   - \mbox{$\frac{p(p-1)}{2}$} \Big [ g_0 \int_{0 < s_1 < t_1 < L_0}
     \langle \Psi_{\alpha \beta} (s_1, t_1) \rangle_0 } 
   \nonumber \\  
   & &+ g_0^2 
     \int_{\stackrel{\mbox{\scriptsize 
         $0 \!<\! s_1 \!<\! t_1 \!<\! s_2 \!<\! t_2
         \!<\! L_0$}}
                    {\mbox{\scriptsize or } 0 < s_1 < s_2 < t_2 < t_1 < L_0}}
     \Big ( \langle \Psi_{\alpha \beta} (s_1, t_1) 
              \Psi_{\alpha \beta} (s_2, t_2) \rangle_0^c 
   \nonumber \\  
   & &   
   + 2 (p-2)
      \langle \Psi_{\alpha \beta} (s_1, t_1) 
              \Psi_{\alpha \gamma} (s_2, t_2) \rangle_0^c 
      \Big )  \Big ]
     + O(g_0^3). 
 \end{eqnarray}
This series contains {\em connected} overlap correlations 
evaluated at $g_0 = 0$. The first-order term involves two, the 
second-order terms involve two and three pairwise different replicas,
respectively; see fig.~2(a)--(c). The integration over the contact points
in (\ref{fpert}) produces a singular dependence of the 
free energy on $g_0$ as well as ultraviolet-divergent terms 
which are regular in $g_0$. Performing these integrals and expanding about the 
point of marginality ($\vep = 0$), we obtain the leading singular part 
\begin{eqnarray} 
\label{fsing}
{\cal F} (u_0, L_0) &=& p
 \Big[ \log L_0 + (p-1) \frac{u_0}{\vep}- \frac{( p-1 ) C_{p}
 u_0^2}{2{\vep }^2}\nonumber  \\
&&\quad  + O \big(\vep, u_0 \vep^0, u_0^2/\vep, u_0^3 \big)
\big]
\end{eqnarray} 
with the dimensionless coupling constant $u_0 = g_0 L_0^{-\vep}$
and $C_p = 1 - 2 (p - 2)$.
\begin{figure}
\fig{.99}{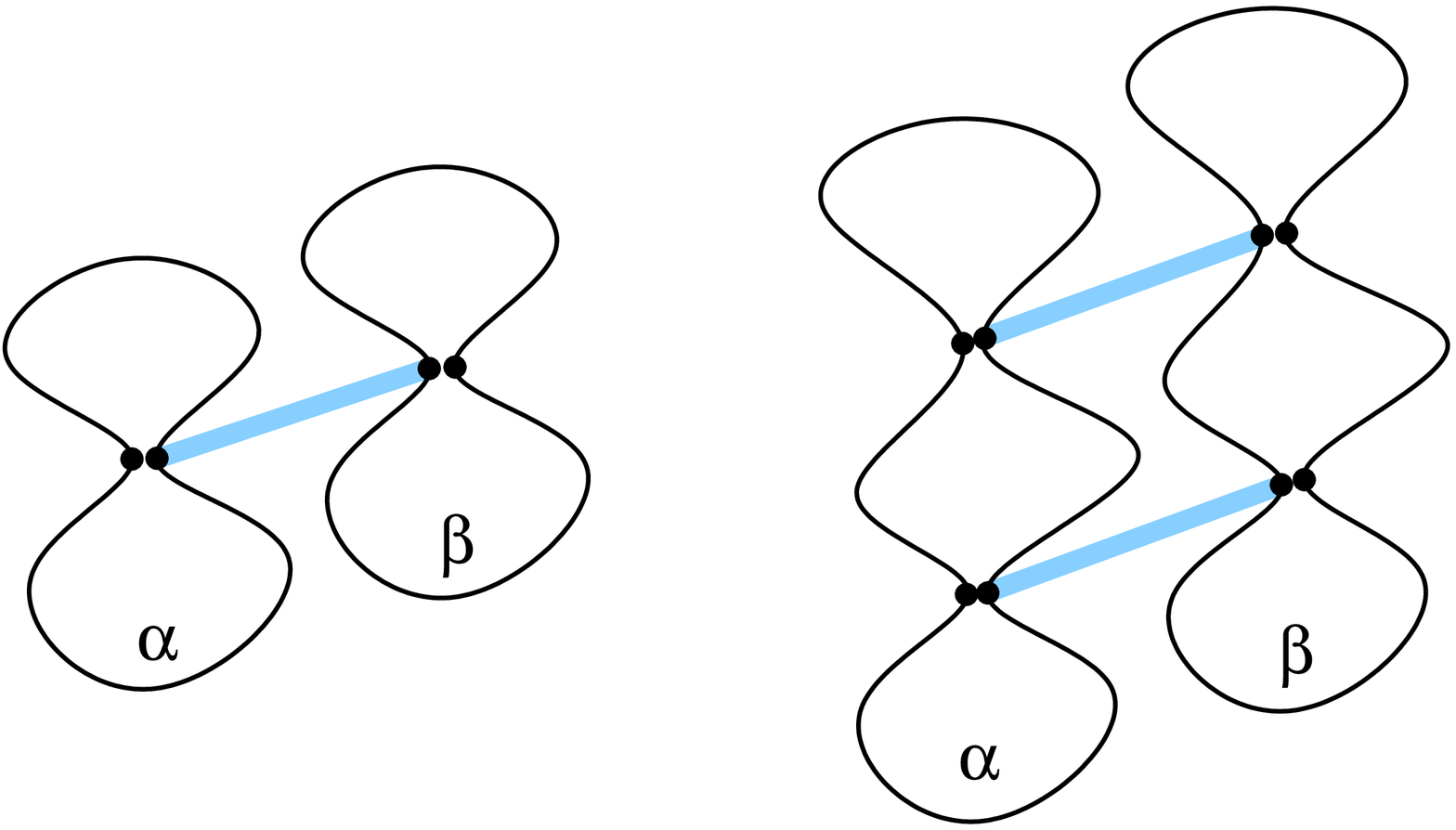}
\vspace*{-0.9cm}

\centerline{\qquad\quad (a) \hfill\qquad ~~~~~(b) \hfill\qquad \qquad
\qquad (c)~~~~} \caption{Overlap correlations in the series (\ref{fpert}).
(a)~Two-replica one-point function $\langle \Psi_{\alpha \beta} (s_1,
t_1) \rangle_0$.  (b)~Two-replica two-point function $\langle
\Psi_{\alpha \beta} (s_1, t_1)
              \Psi_{\alpha \beta} (s_2, t_2) \rangle_0$.
(c)~Three-replica two-point function 
$\langle \Psi_{\alpha \beta} (s_1, t_1) 
              \Psi_{\alpha \gamma} (s_2, t_2) \rangle_0$.
}
\end{figure}
The poles in (\ref{fsing}) are  absorbed into
a renormalized coupling $g = Z_g g_0$ and a renormalized backbone
length $L = Z_L L_0$, such that the free energy becomes an analytic
function of the dimensionless coupling $u = g L^{-\vep}$.  In a
minimal subtraction scheme, we extract from (\ref{fsing}) these
$Z$-factors to leading order,
\begin{equation}
Z_g = 1 -  C_p \frac{u}{\vep} + O(u^2), 
\;
Z_L^{-1} = 1 - (p-1) \frac{u}{\vep} + O(u^2).
\label{Z}
\end{equation}
The resulting renormalization group flow takes a simple form
with respect to the renormalized scale $L$,
\begin{eqnarray}
  \tilde \beta (u) & \equiv & L \frac{\partial}{\partial L} u = -\vep u
  + C_p u^2 + O(u^3), \label{betatilde}
  \\
  \tildegamma (u) & \equiv &
  \frac{L}{L_0} \frac{\partial}{\partial L} L_0 = 1 + (p-1) u 
    + O(u^2).
 \label{gammatilde}
\end{eqnarray}
The beta function is defined 
as the flow with respect to the original scale $L_0$,
\begin{equation}
   \beta (u) \equiv  L_0 \frac{\partial}{\partial L_0} u =
   \frac{\tilde \beta (u)}{\tildegamma (u)} = 
   \frac{-\vep u + C_p u^2 +O(u^3)}{1 + (p-1) u +O(u^2)}.
   \label{beta}
\end{equation} 
It has a nontrivial fixed point $u^* = \vep/ C_p + O(\vep^2)$ for
generic $p$, which is ultraviolet-unstable for $\vep > 0$
and marks the RNA glass transition for $\vep = 1$, $p \to 0$.  The
$\vep$-expansion can be analyzed at higher orders using the {\em
operator product expansion} of the fields $\Phi$ and $\Psi$.
Generalizing the arguments of~\cite{Lassig.KPZ,WieseHabil}, we find
that the theory is renormalizable in $g$ and $L$ (for details,
see~\cite{pre}).  The field $\Phi$ is renormalized by a factor 
$Z_\Phi = Z_L^{-2} + O (u^{2})$~\cite{ao}. By the scaling relation 
$\zeta + \rho = 2$, this 
implies ``superdiffusive'' height fluctuations with exponent $\zeta^*
= \zeta_0/\tildegamma^*+O (\epsilon^{2})$ for $p < 1$, where
$\tildegamma^* \equiv \tildegamma(u^*)$~\cite{ao}.
The renormalization of $\Psi$ is tied to
that of its conjugate coupling $g$.  Hence, the dimensions of $\Phi$
and $\Psi$ at the transition are two independent exponents,
\begin{eqnarray}
\rho^* & = & \frac{\rho_0 + L \partial_L \log Z_\Phi}{\tildegamma (u^*)} =
\frac{1 + \vep/2 + 2 (p-1) \vep /C_p}{1 + (p-1) \vep /C_p} + \dots,
\nonumber
\\
\theta^{*} &=& 2 - \beta ' (u^{*}) =2 -
\frac{\varepsilon}{1+ (p-1)\varepsilon / C_p}+\dots; \qquad  
\label{theta*}
\end{eqnarray}
the omitted terms are of order $p-2$ and $\vep^2$. 
\begin{figure}[t]
\fig{.342}{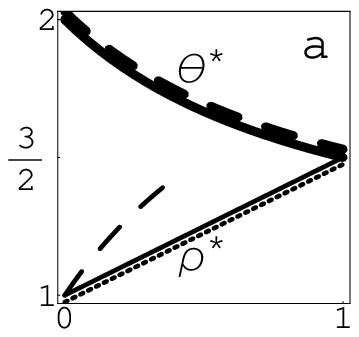} \hspace*{-2.3ex} 
\raisebox{-0ex}{\fig{.325}{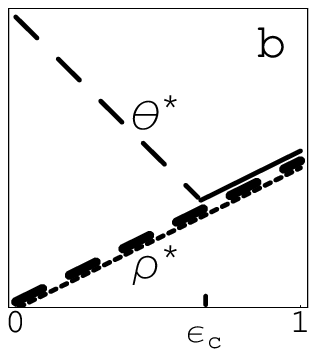} \hspace*{-3.5ex}
\fig{.325}{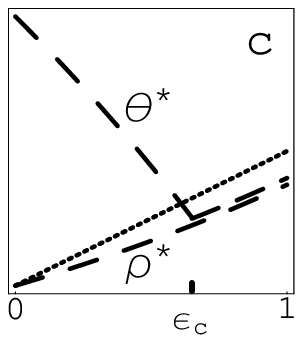}}
\vspace*{-0.2cm}
\caption{The critical exponents $\rho^*$ and $\theta^*$ as a function
of $\vep$ for (a)~$p = 2$, (b) $p = 1$, and (c) $p = 0$. Exact results
(thick solid), renormalization group results valid to all orders (thick
dashed) or to first order (thin dashed), presumably exact values 
(see text, thin solid), reference line $\rho_0 (\vep)$ (dotted).  
}
\end{figure}
These expressions are valid within the constraints $\theta^* \geq \rho^*$,
since two-replica overlap correlations decay at least as fast as single-replica
ones, and $\zeta^* \geq \zeta_0$.  The resulting dependence of the critical
exponents on $p$ and $\epsilon$ is shown in fig.~3.  (a)~For $p=2$, we have
shown that the theory is {\em one-loop renormalizable}, i.e., the expressions
(\ref{Z}) to (\ref{beta}) and (\ref{theta*}) for $\theta^*$ are
exact~\cite{pre}. This reflects the exact summability of the partition function
as shown in~\cite{BH02b} for $\vep = 1$.  We have generalized this solution at
the transition point to arbitrary $\vep$, giving
$\zeta^* = \zeta_0$ and $\rho^* = \rho_0$ (the
renormalization group results are subleading). For $\vep =1$, we thus have
$\theta^* = \rho^* = 3/2$. Hence, two replicas are essentially locked into a
single conformation already at the transition. The borderline value $\vep_c =
1$ corresponds to the upper critical dimension $d_{\mathrm{uc}} = 4$ of
directed polymers \cite{Lassig.KPZ,LassigKinzelbach}.  (b)~For $p = 1$,
renormalization gives $\rho^* = \rho_0$ exactly to all orders, and $\theta^* =
2 - \vep+O (\vep^{2}) $. This produces a borderline value $\vep_c \approx 
2/3$, beyond which $\theta^* = \rho^* = \rho_0$ exactly.  (c)~For $p = 0$, the
first-order eq.~(\ref{theta*}) produces an even smaller value of $\vep_c$. For
$\vep = 1$, we find locked configurations with  $\theta^* = \rho^* = 2 - \zeta^*
\approx 11/8$ as reported above. For $\vep > \vep_c$, the
renormalization-group exponent $\theta^*$ in (\ref{theta*}) describes a {\em
subleading} singularity in the overlap correlations, which is related to rare
critical fluctuations within the locked state~\cite{pre}, cf.~\cite{Lipowsky}
for directed polymers. 

Despite its technical difficulties, our renormalization is rather
intuitive since it acts directly on the fold configurations of
Fig.~1. In a Wilson scheme, we would produce coarse-grained folds with
varying short-distance cutoff $\ell_{\min}$ by integrating out
subconfigurations of backbone length $\ell < \ell_{\min}$.  This leads
to a scale-dependent backbone length $L$ and coupling constant $g$.
For $p > 1$, the attractive replica interaction produces additional
short loops, which are cut off under coarse-graining, i.e., the
effective length is {\em shorter} than without interaction ($L \sim
L_0^{1/\gamma_L^*}$ with $\gamma_L^* > 1$).  For $p < 1$, however,
this effect is reversed ($\gamma_L^* < 1$): $L$ becomes {\em longer}
and the random walk $h(r)$ becomes correlated with superdiffusive
fluctuations ($\zeta^* = 1/2$, $ \gamma_L^* > 1/2$). Hence, the
probability of first return is shifted from small to large scales,
i.e. there are more pairings between distant nucleotides ($\rho^* <
\rho_0$). The locking of pairing correlations ($\theta^* = \rho^*$) at
criticality means that disorder has already its maximal effect on
scaling, i.e., the same exponents govern the glass phase. This
prediction is remarkable in contrast to random directed polymers,
where the roughening transition has no locking for $2 < d < d_{\rm
uc}$ and the low-temperature physics is governed by a new
strong-coupling fixed point.

We thank R. Bundschuh and F. David for discussions.

\vfill


\vspace*{-2ex}
\begin{thebibliography}{99}
\vspace*{-4ex}


\bibitem{Tinoco}
I. Tinoco and C. Bustamante, J. Mol. Biol. 293 (1999), 271. 

\bibitem{Higgs.review}
P.G. Higgs, Quart. Rev. Biophys. 33, 3 (2000). 

\bibitem{McCaskill}
J.S. McCaskill, Biopolymers 29, 1105 (1990).  

\bibitem{HofackerAl}
I.L. Hofacker et al., Monatsh. Chem. 125, 167 (1994). 

\bibitem{Isambert}
H. Isambert and E.D. Siggia, Proc. Natl. Acad. Sci. 97, 6515 (2000).

\bibitem{OrlandZee2002} H.\ Orland, T.\ Zee, Nucl. Phys. B 620, 456
(2002).

\bibitem{deGennes}
P.G. deGennes, Biopolymers 6, 715 (1968). 

\bibitem{Page}
For an introduction, see, R.D. M. Page and E.C.
Holmes, {\em Molecular Evolution}, Blackwell Science, London
(1998). 

\bibitem{seq}
For sequence disorder, the free energy variables factorize,  
$\eta (s,t) = \tilde \eta_1 (s) \tilde \eta_2 (t)$, and are therefore 
correlated. However, treating them as independent random variables 
is numerically an excellent approximation~\cite{BH02b,Krzakala02}. 

\bibitem{BH99}
R. Bundschuh and T. Hwa, Phys. Rev. Lett. 83, 1479 (1999). 

\bibitem{BH02a}
R. Bundschuh and T. Hwa, Europhys. Lett. 59, 903 (2002). 

\bibitem{BH02b}
R. Bundschuh and T. Hwa,
Phys. Rev. E 65, 031903 (2002). 

\bibitem{Krzakala02}
F. Krzakala, M. M\'ezard, and M. M\"uller, 
Europhys. Lett. 57, 752 (2002). 

\bibitem{Higgs96}
P.G. Higgs, Phys. Rev. Lett. 76, 704 (1996). 

\bibitem{Pagnani2000}
A. Pagnani, G. Parisi, and F. Ricci-Tersenghi, Phys. Rev. Lett. 84, 
2026 (2000). 

\bibitem{Hartmann01}
A.K. Hartmann, Phys. Rev. Lett. 86, 1382 (2001). 

\bibitem{Marinari02}
E. Marinari, A. Pagnani, and F. Ricci-Tersenghi,
Phys. Rev. E 65, 041919 (2002).

\bibitem{Liu}
T. Liu and R. Bundschuh, Phys. Rev.  E 69, 061912 (2004). 

\bibitem{HuseHenley}
D.A. Huse and C.L. Henley, Phys. Rev. Lett. 54, 2708 (1985). 

\bibitem{KPZ}
M. Kardar, G. Parisi, and Y.C. Zhang, Phys. Rev. Lett. 56, 889 (1986). 

\bibitem{Lassig.review}
For a review, see, M. L\"assig, J. Phys. C 10, 9905 (1998). 

\bibitem{Lassig.KPZ}
M. L\"assig, Nucl. Phys. B 448, 559 (1995). 

\bibitem{Wiese.KPZ}
K.J. Wiese, J. Stat. Phys. 93, 143-154 (1998).

\bibitem{WieseHabil}
K.J.~Wiese, 
Polymerized Membranes, a Review, 
{\em in} Phase Transitions and Critical Phenomena, vol.~19, C.~Domb
and J.~Lebowitz eds., Acadamic Press, London (2001).

\bibitem{foot1}
Since $\rho^{*}<2$, the integral is convergent and there is no additional
renormalization. 

\bibitem{BL9602}
R. Bundschuh and M. L\"assig, 
Phys. Rev. E 54, 304 (1996) and Phys. Rev. E 65, 061502 (2002).

\bibitem{Bundschuh.pc}
T. Liu and R. Bundschuh, private communication. 

\bibitem{freefield}
Using the height field $h(r)$, secondary structures can be 
mapped  onto {\em free} directed paths ${\bf h} (r) \in
\mathbb{R}^d$ with 
$d = 3$~\cite{BH99}. In this representation, the
analytic continuation takes place in the dimension
$d = 2 \rho_0$.   


\bibitem{pre}
M. L\"assig and K.J. Wiese, to be published. 

\bibitem{ao}
We conjecture that this identity is valid to all orders.

\bibitem{LassigKinzelbach}
M. L\"assig and H. Kinzelbach, Phys. Rev. Lett. 78, 903 (1997). 

\bibitem{Lipowsky}
R. Lipowsky, Europhys. Lett. 15, 703 (1991). 



\end{thebibliography}
\end{document}